\documentclass[twocolumn,showpacs,aps,prl,letterpaper]{revtex4}
\usepackage{graphicx}
\usepackage{dcolumn}
\usepackage{amsmath}
\usepackage{epsfig}

\long\def\inst#1{\par\nobreak\kern 4pt\nobreak
    {\itshape #1}\par\vskip 10pt plus 3pt minus 3pt}
\RequirePackage{xspace}

\def\babar{\mbox{\slshape B\kern-0.1em{\smaller A}\kern-0.1em
    B\kern-0.1em{\smaller A\kern-0.2em R}}}
\def\Kbar    {\kern 0.18em\overline{\kern -0.18em K}{}\xspace}

\def\Kz      {\ensuremath{K^0}\xspace}
\def\Kzb     {\ensuremath{\Kbar^0}\xspace}
\def\KzKzb   {\ensuremath{\Kz {\kern -0.16em \Kzb}}\xspace}

\def\Ks     {\ensuremath{K_S}\xspace}
\def\Kl     {\ensuremath{K_L}\xspace}
\def\KsKs   {\ensuremath{\Ks {\kern -0.16em \Ks}}\xspace}
\def\KlKl   {\ensuremath{\Kl {\kern -0.16em \Kl}}\xspace}
\def\KsKl   {\ensuremath{\Ks {\kern -0.16em \Kl}}\xspace}
\def\KlKs   {\ensuremath{\Kl {\kern -0.16em \Ks}}\xspace}
\def\Dbar    {\kern 0.18em\overline{\kern -0.18em D}{}\xspace}
\def\Dz      {\ensuremath{D^0}\xspace}
\def\Dzb     {\ensuremath{\Dbar^0}\xspace}
\def\DzDzb   {\ensuremath{\Dz {\kern -0.16em \Dzb}}\xspace}
\def\Bbar    {\kern 0.18em\overline{\kern -0.18em B}{}\xspace}

\def\Bz      {\ensuremath{B^0}\xspace}
\def\Bzb     {\ensuremath{\Bbar^0}\xspace}
\def\BzBzb   {\ensuremath{\Bz {\kern -0.16em \Bzb}}\xspace}
\def\Bu      {\ensuremath{B^+}\xspace}
\def\Bub     {\ensuremath{B^-}\xspace}

\def\BpBm    {\ensuremath{\Bu {\kern -0.16em \Bub}}\xspace}

\newcommand{\optbar}[1]{\shortstack{{\tiny (\rule[.4ex]{1em}{.1mm})}
  \\ [-.7ex] $#1$}}
\def\BorBbar    {\kern 0.18em\optbar{\kern -0.18em B}{}\xspace}
\def\DorDbar    {\kern 0.18em\optbar{\kern -0.18em D}{}\xspace}
\def\KorKbar    {\kern 0.18em\optbar{\kern -0.18em K}{}\xspace}

\def\pep2{PEP-II}
\mathchardef\Upsilon="7107
\def\Y#1S{\ensuremath{\Upsilon{(#1S)}}\xspace}

\begin{document}

\title{\large \bfseries \boldmath Rare Semileptonic Decays of Heavy Mesons
with Flavor SU(3) Symmetry}
\author{Hai-Bo Li$^2$}\email{lihb@ihep.ac.cn}
\author{Mao-Zhi Yang$^1$}\email{yangmz@nankai.edu.cn}
\affiliation{$^1$ Department of Physics, Nankai University,
Tianjin, 300071, China \\$^2$Institute of High Energy Physics,
P.O.Box 918, Beijing  100049, China}


\date{\today}


\begin{abstract}
In this paper, we calculate the decay rates of $D^+ \rightarrow
D^0 e^+ \nu$, $D^+_S \rightarrow D^0 e^+ \nu$, $B^0_S \rightarrow
B^+ e^- \overline{\nu}$, $D^+_S \rightarrow D^+ e^- e^+$ and
$B^0_S \rightarrow B^0 e^-e^+$ semileptonic decay processes, in
which only the light quarks decay, while the heavy flavors remain
unchanged. The branching ratios of these decay processes are
calculated with the flavor SU(3) symmetry. The uncertainties are
estimated by considering the SU(3) breaking effect. We find that
the decay rates are very tiny in the framework of the Standard
Model. We also estimate the sensitivities of the measurements of
these rare decays at the future experiments, such as BES-III,
super-$B$ and LHC-$b$.

\end{abstract}


\maketitle

The properties of hadrons containing a single heavy quark Q ($m_Q
\gg \Lambda_{QCD}$) along with light degrees of freedom are
constrained by symmetries which are not apparent in QCD~\cite{1}.
These symmetries are manifest in the heavy quark effective theory
(HQET) where the heavy quark acts in the hadron's rest frame like
a spatially static triplet source of color electric
field~\cite{2,neubert96}. In the HQET the heavy quark's coupling
to the gluon degrees of freedom are independent of its mass and
described by a Wilson line~\cite{3}. In the rare decay processes
of $D^+ \rightarrow D^0 e^+ \nu$, $D^+_S \rightarrow D^0 e^+ \nu$,
$B^0_S \rightarrow B^+ e^- \overline{\nu}$, $D^+_S \rightarrow D^+
e^- e^+$ and $B^0_S \rightarrow B^0 e^-e^+$, the heavy quark
flavors ($c$ or $b$) remain unchanged, and the weak decays are
managed by the light quark sectors.  In the limit of the flavor
SU(3) symmetry of the light quarks, the matrix elements of the
weak current can be constrained, the uncertainty can be estimated.

In this work, we study the rare decay processes $D^+ \rightarrow
D^0 e^+ \nu$, $D^+_S \rightarrow D^0 e^+ \nu$, $B^0_S \rightarrow
B^+ e^- \overline{\nu}$, $D^+_S \rightarrow D^+ e^- e^+$ and
$B^0_S \rightarrow B^0 e^-e^+$. Applying the SU(3) symmetry for
the light quarks, the form factors describing the strong
interaction in these decays can be obtained. The uncertainties can
be estimated by considering SU(3) breaking effect.

For the semileptonic decays $D^+ \rightarrow D^0 e^+ \nu$ and
$D^+_S \rightarrow D^0 e^+ \nu$, the decay amplitude is
\begin{eqnarray}
 {\cal A} = && \frac{G_F}{\sqrt{2}} V_{ij} \Bigl[\bar{u}(k_1)
 \gamma^{\mu} (1-\gamma_5) v(k_2) \nonumber \\
 && \langle D^0 (p_2)| \bar{q}_1
 \gamma_{\mu}(1-\gamma_5) q_2| D^+_{(S)} (p_1) \rangle \Bigr],
\label{eq:semi-amp}
\end{eqnarray}
where $G_F= 1.16639 \times 10^{-5} \, \mbox{GeV}^{-2}$ is the
Fermi constant,
and $V_{ij}$ is the Cabibbo-Kobayashi-Maskawa (CKM) matrix
element, the functions $\bar{u}(k_1)$ and $v(k_2)$ are Dirac
spinors, which describe the productions of the neutrino with
momentum $k_1$ and the anti-lepton with momentum $k_2$,
respectively.

According to its Lorentz structure, the hadronic matrix element in
eq.(\ref{eq:semi-amp}) can be decomposed as
\begin{eqnarray}
&&\langle D^0 (p_2)| \bar{q}_1
 \gamma_{\mu}(1-\gamma_5) q_2| D^+_{(S)} (p_1) \rangle\nonumber \\& = & f_+(q^2)
 (p_1+p_2)_{\mu} +
 f_-(q^2)(p_1-p_2)_{\mu},
 \label{eq:define}
\end{eqnarray}
where $f_{\pm}(q^2)$ are the form factors including all the
dynamics of strong interaction. Considering the decomposition of
the hadronic matrix element in eq.(\ref{eq:define}), and
neglecting the lepton mass, we get the amplitude as
\begin{eqnarray}
 {\cal A} = \frac{G_F}{\sqrt{2}} V_{ij} f_+(q^2)(p_1+p_2)_{\mu}
 \bar{u}(k_1) \gamma^{\mu} (1-\gamma_5) v(k_2).
 \label{eq:define2}
\end{eqnarray}
Then the decay width can be obtained as
\begin{eqnarray}
 \Gamma &=& \frac{1}{192 \pi^3 m_1^3} G^2_F |V_{ij}|^2 \times \nonumber \\
         && \int dq^2
 f^2_+(q^2)\bigl[(m_1^2+m^2_2 -q^2)^2 - 4m_1^2m^2_2\bigr]^{3/2},
\label{eq:width}
\end{eqnarray}
where $m_1$ and $m_2$ are the masses of the initial and final
heavy hadron involved in these rare decays.

Next we consider the flavor-changing neutral current (FCNC)
processes $D^+_S \rightarrow D^+ e^- e^+$ and $B^0_S \rightarrow
B^0 e^- e^+$. They can be described by the $\Delta S = 1$
effective Hamiltonian in the quark level at scales $\mu < m_c$
\cite{4}:
\begin{eqnarray}
{\cal H}_{eff}^{\Delta S = 1} &=& \frac{G_F}{\sqrt{2}} \Bigl[
\sum^{6, 7V}_{i=1} (V^*_{us}V_{ud} Z_i(\mu)-V^*_{ts}V_{td}
Y_i(\mu))Q_i(\mu) \nonumber \\
&&- V^*_{ts}V_{td} Y_{7A}(m_W)Q_{7A}(m_W) \Bigr],
 \label{eq:time:amplitude_s}
\end{eqnarray}
where the operators $Q_i(\mu)$'s are defined as
\begin{eqnarray}
 Q_1 &=& (\bar{s}_\alpha u_\beta)_{V-A} (\bar{u}_\beta d_\alpha)_{V-A} \nonumber \\
 Q_2 &=& (\bar{s}u)_{V-A}(\bar{u}d)_{V-A} \nonumber \\
 Q_3 &=& (\bar{s}d)_{V-A} \sum_{q} (\bar{q}q)_{V-A} \nonumber \\
 Q_4 &=& (\bar{s}_\alpha d_\beta)_{V-A} \sum_q (\bar{q}_\beta q_\alpha)_{V-A}
 \nonumber \\
 Q_5 &=& (\bar{s}d)_{V-A} \sum_{q}(\bar{q}q)_{V+A} \nonumber \\
 Q_6 &=& (\bar{s}_\alpha d_\beta)_{V-A} \sum_q (\bar{q}_\beta q_\alpha)_{V+A}
 \nonumber \\
 Q_7 &=& (\bar{s}d)_{V-A} (\bar{e}e)_V \nonumber \\
 Q_{7A} &=& (\bar{s}d)_{V-A} (\bar{e}e)_A.
 \label{eq:wilson_coeff}
\end{eqnarray}
Here the indexes $\alpha$ and $\beta$ are color numbers, the
summation $\sum$ runs over all the quark flavors which are active
at the scale $\mu$, and the $V\pm A$ denotes $\gamma_\mu (1\pm
\gamma_5)$.

 At $\mu =1 $ GeV with $\Lambda^{(4)}_{\bar{\mu s}} = 215$ MeV
and NDR scheme, the Wilson coefficients $Z_i$ and $Y_i$'s are
calculated to be \cite{4} $Z_1 = -0.409$, $Z_2 = 1.212$, $Z_3 =
0.008$, $Z_4= -0.022$, $Z_5 = 0.006$, $Z_6= -0.022$, $Z_{7V} =
-0.015\alpha_{QED}$, and $Y_1=Y_2=0$, $Y_3=0.025$, $Y_4 = -0.048$,
$Y_5=0.005$, $Y_6 = -0.078$, $Y_{7V} = 0.747 \alpha_{QED}$,
$Y_{7A} = -0.700 \alpha_{QED}$, where $\alpha_{QED}$ is the fine
structure constant.

\begin{figure}[h]
\epsfig{file=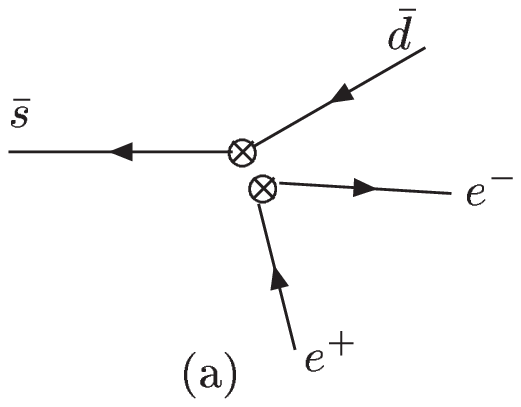,width=4cm,height=3.5cm}
\epsfig{file=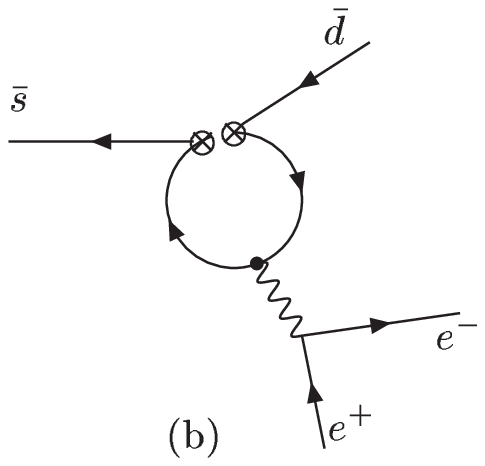,width=4cm,height=4.0cm}
\epsfig{file=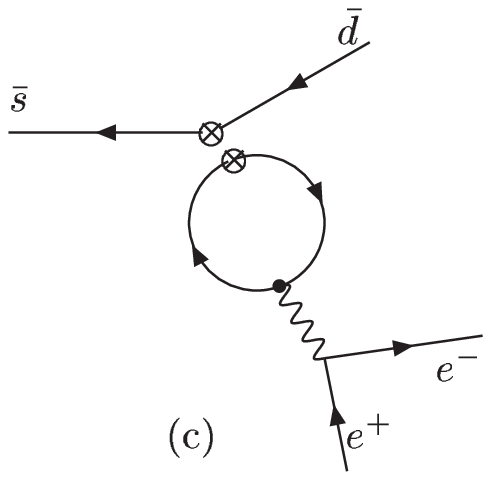,width=4cm,height=4.0cm} \caption{Diagrams
for the rare decays of $D^+_S \rightarrow D^+ e^+ e^-$ and
$\bar{B}^0_S \rightarrow \bar{B}^0 e^+ e^-$ up to one-loop level,
where the cross circles denote the operator insertion. (a): the
tree level diagram; (b) and (c): the one-loop level diagrams with
two different ways of operator insertions. } \label{treeloop}
\end{figure}

We calculate the FCNC processes $D^+_S \rightarrow D^+ e^+ e^-$
and $\bar{B}^0_S \rightarrow B^0 e^+ e^-$ up to one-loop level
based on the effective Hamiltonian. The diagrams we considered are
depicted in Fig.\ref{treeloop}. The amplitude for the FCNC
processes is calculated to be
\begin{eqnarray}
{\cal A} &= &\bigl[  a_1 \bar{u}(k_1)\gamma^{\mu} v(k_2)+
a_2\bar{u}(k_1)\gamma^{\mu}\gamma_5
v(k_2)\bigr]\nonumber \\
&& \times (p_1+p_2)_{\mu}f_+(q^2),
 \label{eq:ampl_fcnc}
\end{eqnarray}
with
\begin{eqnarray}
a_1 &=& \frac{G_F}{\sqrt{2}}V^*_{us}V_{ud}\Bigl\{Z_{7V} + \tau
Y_{7V} +
\frac{N_c\alpha_{QED}}{2\pi}\times \nonumber \\
&&\Bigl[\bigl((Z_1+\frac{Z_2}{N_c})Q_u + (\frac{Z_3}{N_c}+Z_4)Q_d
+
\tau(\frac{Y_3}{N_c}+Y_4)Q_d\bigr)\times \nonumber \\
&&\bigl(-\frac{2}{3}+G(0,q^2,\mu)\bigr)+\bigl(Z_3+\frac{Z_4}{N_c}+Z_5+\frac{Z_6}{Nc}+\nonumber
\\ &&\tau(Y_3+\frac{Y_4}{N_c}+Y_5+\frac{Y_6}{N_c})\bigr)\bigl((Q_u+Q_d)G(0,q^2,\mu)+\nonumber \\
&&Q_sG(m^2_s,q^2,\mu)\bigr)+ (Z_{7V}+\tau
Y_{7V})\frac{Q_e}{N_c}G(0,q^2,\mu) \Bigr]\Bigr\}, \label{eq:a1}\\
 a_2 &=&
\frac{G_F}{\sqrt{2}}V^*_{us}V_{ud} \tau Y_{7A},
 \label{eq:a2}
\end{eqnarray}
where $N_c=3$ is the color number, $Q_u$, $Q_d$, $Q_s$ and $Q_e$
are the charge of the relevant quarks and electrons, $m_s$ is the
mass of the strange quark, and the function $G$ is defined as
\begin{eqnarray}
G(m^2,q^2,\mu) &=& -\int^1_0 dx 4x (1-x)\times \nonumber \\
 &&\ln\frac{m^2-x(1-x)q^2-i \epsilon}{\mu^2},
 \label{eq:g_define}
\end{eqnarray}
which is originated from the loop calculation.  The parameter
$\tau$ is defined to be
\begin{eqnarray}
\tau = -\frac{V^*_{ts}V_{td}}{V^*_{us}V_{ud}}.
 \label{eq:ckm_ratio}
\end{eqnarray}
With the amplitude in eq.(\ref{eq:ampl_fcnc}), the decay width can
be obtained as
\begin{eqnarray}
 \Gamma &=& \frac{1}{192 \pi^3m^3_1} \int dq^2 f_+(q^2)^2(|a_1|^2
 +|a_2|^2)\times \nonumber \\
&& \bigl[(m_1^2+m_2^2 - q^2)^2-4m^2_1 m^2_2 \bigr]^{3/2}.
 \label{eq:decay_fcnc}
\end{eqnarray}
In the numerical calculations we use the following values for the
Wolfenstein parameters \cite{PDG2006}:
\begin{eqnarray}
\lambda = 0.2272; \,\, A=0.818; \,\, \bar{\rho}=0.221;\,\,
\bar{\eta} = 0.340.
 \label{eq:ckm_par}
\end{eqnarray}

In the limit of SU(3) symmetry for the light quarks $u$, $d$ and
$s$, the masses of the light quarks are the same, the masses of
the mesons with the same heavy quark but different light flavors
are also the same. Therefore the energy release in the rare decay
process is zero in the limit of the SU(3) symmetry. The velocity
of the heavy mesons in the initial and final states is unchanged.
The wave functions of the light quarks are the same after the rare
decay occurs. The heavy quark does not feel any change during the
light flavor transition process. Thus with the flavor SU(3)
symmetry holding, the form factor $f_+(q^2)$ can be normalized to
unity at the point $q^2=0$,
\begin{equation}
f_+(0) =1. \label{SU3}
\end{equation}
Note that there is only one kinematic value for $q^2$ with the
SU(3) symmetry holding, $q^2=0$. However, with SU(3) flavor
symmetry broken, i.e., the mass of $s$ quark $m_s$ being larger
than that of $u$ and $d$ quarks, the light quark is boosted after
the light flavor transition occurs. The possibility for the light
and heavy quarks still bound together to form a heavy meson
decreases, this leads to a form-factor suppression. Thus the value
of the form factors must be smaller than unity. The larger the $s$
quark mass, the larger the deviation.

With the SU(3) symmetry broken, the allowed value of $q^2$ is
extended from $q^2=0$ to a range of $0\le q^2\le q^2_{max}$. At
the point $q^2=q^2_{max}$, all the energy and momentum released in
the decay are carried away by the lepton pair, the velocity of the
final meson is not changed (this is the point of zero recoil) ,
only the mass of the light quark changed, therefore the deviation
from the limit of SU(3) symmetry is the smallest at the point of
zero recoil. For the point $q^2=0$, there is further deviation
caused by the kinematic change of the final meson. Thus one can
write the the form factor at $q^2=q^2_{max}$ as an expansion in
terms of a small SU(3) breaking parameter. Including the SU(3)
breaking effect, eq.(\ref{SU3}) is extended to the point at
$q^2=q^2_{max}$
\begin{equation}
f_+(q^2_{max}) =1+\lambda_{SU(3)},  \label{SU3B}
\end{equation}
where the parameter $\lambda_{SU(3)}$ describes the correction due
to SU(3) breaking effect at $q^2=q^2_{max}$. Note that the value
of $\lambda_{SU(3)}$ must be negative, because the SU(3) breaking
effect leads to a form-factor suppression. In addition, if the
expansion including the SU(3) breaking effect is performed at the
point $q^2=0$ instead of the point of zero recoil, further
correction due to the velocity change of the final meson should be
considered.

The larger the mass of the $s$ quark (here we neglect the masses
of $u$ and $d$ quarks), the larger the SU(3) breaking effects. An
appropriate SU(3) breaking parameter can be taken to be
$m_s/\Lambda$, where $\Lambda \sim 1\mathrm{GeV}$ is a hadronic
scale. According to the Ademollo-Gatto theorem
\cite{AGtheorem,AG2}, the parameter $\lambda_{SU(3)}$ should be of
the second order of the SU(3) breaking parameter, i.e.,
\begin{equation}
\lambda_{SU(3)} \sim {\cal{O}}((m_s/\Lambda)^2).
\end{equation}
In this work the correction to the form factor due to the SU(3)
breaking effect is treated as an uncertainty, the form factor is
varied in the range
\begin{equation}
1-(m_s/\Lambda)^2\le f_+(q^2_{max})\le 1.\label{r1}
\end{equation}

The momentum transfer squared $q^2$ should be in the range range
$0< q^2<(m_{H_1}-m_{H_2})^2$, where $m_{H_1}$ and $m_{H_2}$ are
the masses of the initial and final mesons containing the heavy
quark, respectively. The $q^2$ dependence of the form factor is
typically governed by the nearest resonance, which can be
parameterized as a pole-dominance form \cite{pole}
\begin{equation}
f_+(q^2)=\frac{f_+(0)}{1-q^2/m^2_{H^*}},\label{q-sqare}
\end{equation}
where $H^*$ is nearest pole resonance which contributes to the
from factor. $H^*$ should be $K^*$ in $s\to d$, or $u$
transitions. With eqs.(\ref{r1}) and (\ref{q-sqare}), we can
obtain that the form factor at $q^2=0$ should be varied in the
range
\begin{equation}
(1-(m_s/\Lambda)^2)(1-q^2_{max}/m^2_{H^*})\le f_+(0)\le
(1-q^2_{max}/m^2_{H^*}).\label{r2}
\end{equation}

Taking the mass of strange quark $m_s = 100$ MeV, by considering
the SU(3) breaking effect, we get the following branching
fractions:
\begin{eqnarray}
{\cal B}(D^+ \rightarrow D^0 e^+\nu) &= &2.78 \times 10^{-13}, \nonumber \\
{\cal B}(D^+_S \rightarrow D^0 e^+\nu) &= &(2.97\pm 0.03)\times 10^{-8}, \nonumber \\
{\cal B}(B^0_S \rightarrow B^+ e^- \bar{\nu}) &=& (4.46\pm 0.05) \times 10^{-8},\nonumber \\
{\cal B}(D^+_S \rightarrow D^+ e^+ e^-) &= & (3.56\pm 0.04) \times 10^{-17}, \nonumber \\
{\cal B}(B^0_S \rightarrow B^0 e^+ e^-) &= & (6.44\pm 0.07)\times
10^{-17},
 \label{eq:bra}
\end{eqnarray}
where the uncertainties are caused by SU(3) breaking effect. For
$D^+ \rightarrow D^0 e^+\nu$, the correction to the form factor
due to isospin breaking effect is neglected in the above
presentation, the numerical correction due to isospin breaking
should be far less than the effects in the strangeness-violating
processes. Here we take $m_c=1.5$ GeV, $m_b=4.8$ GeV in the
numerical calculations.

The mass of the strange quark is very close to the mass
differences of the initial and final mesons in the decay processes
involving $s$ quark, $D^+_S \rightarrow D^0 e^+\nu$, $D^+_S
\rightarrow D^+ e^+ e^-$, $B^0_S \rightarrow B^+ e^- \bar{\nu}$,
$B^0_S \rightarrow B^0 e^+ e^-$, which is about $100$ MeV. It is
interesting to study the role of $m_s$ in these decay processes by
slightly varying the mass of strange quark. The results are given
in Table \ref{table1}.  The decay rates are only slightly changed
with $m_s$ varying in the range
$100~\mbox{MeV}<m_s<140~\mbox{MeV}$.

\begin{table*}[htbp]
\caption{ The decay rates of the rare processes involving strange
quark. The second uncertainty is estimated from SU(3) breaking
effect due to the mass of the strange quark.}\label{table1}
\begin{tabular}{c|c|c|c}\hline
   & $m_s=100$MeV & $m_s=120$MeV &  $m_s=140$MeV \\ \hline
$D^+_S \rightarrow D^0 e^+\nu$& $(2.97\pm 0.03)\times 10^{-8} $ &
$(2.96\pm 0.05)\times 10^{-8}  $& $ (2.94\pm 0.06)\times 10^{-8} $ \\
\hline $B^0_S \rightarrow B^+ e^- \bar{\nu}$ & $(4.46\pm
0.05)\times 10^{-8} $&$(4.43\pm 0.07)\times 10^{-8}$& $(4.41\pm
0.09)\times 10^{-8}$ \\ \hline $D^+_S \rightarrow D^+ e^+ e^-$&
$(3.56\pm 0.04) \times 10^{-17}$ &$(3.64\pm 0.06) \times
10^{-17}$& $(3.63\pm 0.07) \times 10^{-17}$ \\ \hline $B^0_S
\rightarrow B^0 e^+ e^-$& $(6.44\pm 0.07)\times
10^{-17}$&$(6.40\pm 0.10)\times 10^{-17}$&$(6.37\pm 0.13)\times
10^{-17}$\\ \hline
\end{tabular}
\end{table*}

Within the SM framework, we find that the branching fractions for
these rare decays are tiny.  However, in the coming experiments at
BES-III~\cite{bes3_deg}, LHC-$b$~\cite{lhcb} and super-$B$
factory\cite{superb}, it is interesting to search for these
semileptonic decays. Especially, the decays of $ D^+_S \rightarrow
D^0 e^+\nu$ and $B^0_S \rightarrow B^+ e^- \bar{\nu}$ may be
reached at the super-$B$ factory.

Since the electron is very soft, one cannot reconstruct both the
electron and neutrino in the experiment near the charm meson
threshold at $e^+e^-$ colliders. At BES-III, to search for the
decay $D^+ \rightarrow D^0 e^+\nu$ on the $\psi(3770)$ peak, the
charged $D$ mesons are produced in pairs, $e^+ e^- \rightarrow
\psi(3770) \rightarrow D^+ D^-$. Thus, the following six tag
modes, $D^- \rightarrow K^+ \pi^- \pi^-$, $K^+ \pi^-\pi^-\pi^0$, $
K_S \pi^-$, $K_S \pi^- \pi^- \pi^+$, $K_S \pi^- \pi^0$ and
$K^+K^-\pi^-$,  can be used to fully reconstruct one of the
charged $D$ mesons. The summed branching fractions of the six tag
modes are about 28\% of all the charged $D$ decays~\cite{PDG2006}.
The tag efficiency for the charged $D$ mesons are about 20\%,
which means that 20\% of all the $D^+D^-$ pairs can be
tagged~\cite{bes3_deg}. For this case, in order to detect the
decay $D^+ \rightarrow D^0 e^+\nu$, one can reconstruct the
neutral $D$ meson decay by using 46\% of all of the $D^0$ decays
modes in the tagged charged $D$ sample~\cite{bes3_deg}. If we see
any event of the production of the neutral $D$ mesons against the
charged $D$ mesons, it indicates the observation of the rare
semileptonic decay. For the decay $D^+_S \rightarrow D^0 e^+\nu$,
the same method can be applied by using the data collected at the
center of mass $E_{CM} = 4170$ MeV. With 20 fb$^{-1}$ data on the
$\psi(3770)$ peak, the sensitivity of the measurement of $D^+
\rightarrow D^0 e^+\nu$ can be $10^{-6}$ at the BES-III
experiment, while, for the measurement of $D^+_S \rightarrow D^0
e^+\nu$, it can reach $10^{-5}$ level with 20 fb$^{-1}$ data
running at $E_{CM} = 4170$ MeV. These estimations are listed in
table~\ref{tab:experiments},

These semi-leptonic decays can also be searched in the $B$ and
super-$B$ factories by using data at $\Upsilon(4S)$ peak.  One can
reconstruct the following decay chain to search for the rare
decays $D^+ \rightarrow D^0 e^+ \nu$:
\begin{eqnarray}
 D^{*+} \rightarrow D^+ \pi^0_{soft}, \,\,\, D^+ \rightarrow D^0 e^+_{soft} \nu,
 \label{eq:decay_chain}
\end{eqnarray}
where the $D^{*+}$ is boosted, and both $\pi^0$ and electron could
have momentum with a few hundred MeV, which can be detected and
reconstructed in the detector. Since the missing neutrino has very
low momentum, one can partially reconstruct the decay of $D^+$,
and looking at the mass difference $\Delta m = m((D^0 e^+)
\pi^0_{soft}) - m(D^0 e^+)$. The signal events should peak around
the mass difference of $m_{D^{*+}} -m_{D^+}= 140$ MeV on the
$\Delta m$ distribution. In the mass difference, the uncertainty
of the reconstruction of $(D^0 e^+)$ can cancel. The resolution on
the $\Delta m$ will be dominated by the detection of the soft
$\pi^0$. This is a powerful variable to separate background from
the signal events. For the decay $D^+_S \rightarrow D^0 e^+\nu$,
one can use the reaction:
\begin{eqnarray}
 D^{*+}_S \rightarrow D^+_S \gamma_{soft}, \,\,\, D^+_S \rightarrow D^0 e^+_{soft}
 \nu,
 \label{eq:decay_chain_Ds}
\end{eqnarray}
to extract the rare decay signal by looking at the mass difference
$\Delta m = m((D^0 e^+) \gamma_{soft}) - m((D^0 e^+)$. With 1
ab$^{-1}$ and 50 ab$^{-1}$ luminosity at $B$ factories and
super-$B$, the sensitivities could be $10^{-8}$ and $10^{-10}$
respectively. In table~\ref{tab:experiments}, the sensitiveities
of the measurements of the rare $D^+$ and $D^+_S$ decays are
summarized at different experiments.

At super-$B$ factory,  the data taken at $\Upsilon(5S)$ can be
used to search for the rare decays $B^0_S \rightarrow B^+ e^-
\bar{\nu}$ and $B^0_S \rightarrow B^0 e^- e^+$. The cross section
of the $\Upsilon(5S)$ production at $e^+e^-$ collisions is
$\sigma(e^+e^- \rightarrow \Upsilon(5S)) = 0.301 \pm 0.002\pm
0.039 $ nb~\cite{cleo-5s}. Unlike the $\Upsilon(4S)$ state,
$\Upsilon(5S)$ is heavy enough to decay into several $B$ meson
states, in which the vector-vector ($B^*_S \bar{B}^*_S$) and
vector-pseudoscalar ($B^*_S \bar{B}_S + B_S \bar{B}^*_S$)
combinations are dominant~\cite{italian}.  About 30\% of the
$\Upsilon(5S)$ decays into $B_S$ final states~\cite{superb}. With
30 ab$^{-1}$ data at $\Upsilon(5S)$ peak at super-$B$, the
sensitivity of the measurements of the rare $B_S$ decays can be
$10^{-9}$ by assuming 30\% efficiency for the $B_S$
reconstruction.
\begin{table}[htbp]
  \centering
  \caption{ Experimental sensitivities at BES-III, $B$ factory,
  Super-$B$ and LHC-$b$ for the rare decays. We assume the integrated
  luminosities are 20 fb$^{-1}$ (BES-III at $\psi(3770)$ peak
   and 4170 MeV),
  1 ab$^{-1}$ and 50 ab$^{-1}$ at $B$ factory and Super-$B$
  (at $\Upsilon(4S)$ peak), 10 fb$^{-1}$ at LHC-$b$, respectively. }
  \begin{tabular}{c|c|c|c|c} \hline\hline
  Decays & BES-III  & $B$ factory & Super-$B$ & LHC-$b$\\
         & ($\times 10^{-6}$) & ($\times 10^{-8}$) & ($\times 10^{-10}$) & ($\times 10^{-9}$)\\ \hline
  $D^+ \rightarrow D^0 e^+\nu$ & 1.0 & 1.1 &  2.3&  3.8 \\
  $D^+_S \rightarrow D^0 e^+ \nu$ & 5.0 & 1.1 &  2.3 &  3.8 \\
  $D^+_S \rightarrow D^+ e^+ e^- $ & 11.5 & 2.0 & 4.6 &  5.0 \\
  \hline \hline
 \end{tabular}
  \label{tab:experiments}
\end{table}

All of these rare decays can be studied at the LHC-$b$, in which
production cross section for $b\bar{b}$ is 500 $\mu b$. About $5
\times 10^{12}$ $b\bar{b}$ pairs will be obtained in 10 fb$^{-1}$
integrated luminosity in the first five years running at the
LHC-$b$~\cite{lhcb}. Accordingly, about 40\% (40\% or 10\%) of the
$b\bar{b}$ pairs is predicted to form $B^0$ meson ($B^+$ or
$B^0_S$ meson). Thus, the sensitivity of the measurements of the
rare $B_S$ decays is estimated to be $10^{-10}$ at the LHC-$b$
with 10 fb$^{-1}$ luminosity. Both the $D^{*+}$ and $D^{*+}_S$ can
be reconstructed in the decays of $B$ and $B^0_S$ mesons, the
decay chains in Eqs. \ref{eq:decay_chain} and
\ref{eq:decay_chain_Ds} can also be used to extract the rare decay
signals of $D^+$ and $D^+_S$ mesons.  The estimated sensitivities
for the rare decays of the charm mesons are about $10^{-9}$ at the
LHC-$b$. However, one has to be careful that there are many
unexpected backgrounds at the LHC-$b$ experiment. The estimated
sensitivity should be more conservative.

In summary, for the first time, we calculated the decay rates of
the rare $D^+$, $D^+_S$ and $B_S$ decays, in which only the light
quarks decay weakly, while the heavy flavors remain unchanged.
Applying the SU(3) flavor symmetry, the form factors describing
the strong interaction in these decays can be obtained.
Considering SU(3) symmetry breaking, the uncertainty for the form
factors can be estimated. Therefore, these rare decays can be
predicted with the uncertainty estimated by considering SU(3)
symmetry breaking. We also estimated the sensitivities of the
measurements of these rare decays at the future experiments, such
as BES-III, super-$B$ and LHC-$b$. Especially, the decays of $
D^+_S \rightarrow D^0 e^+\nu$ and $B^0_S \rightarrow B^+ e^-
\bar{\nu}$ may be reached at the super-B factory.  Observations of
these decays will be used to test the SM predictions for the rare
simleptonic decays. Further more, any indication of deviation from
the SM prediction may shed light on the searches of New Physics.

The authors thank M.B. Voloshin for useful comments. This work is
supported in part by the National Natural Science Foundation of
China under contracts Nos. 10575108,10521003, 10735080, the 100
Talents program of CAS, and the Knowledge Innovation Project of
CAS under contract Nos. U-612 and U-530 (IHEP).




\end{document}